\documentclass{article}

\PassOptionsToPackage{numbers, compress}{natbib}
\usepackage[preprint]{neurips_2025}

\usepackage[utf8]{inputenc} % allow utf-8 input
\usepackage[T1]{fontenc}    % use 8-bit T1 fonts
\usepackage{hyperref}       % hyperlinks
\usepackage{url}            % simple URL typesetting
\usepackage{booktabs}       % professional-quality tables
\usepackage{amsfonts}       % blackboard math symbols
\usepackage{nicefrac}       % compact symbols for 1/2, etc.
\usepackage{microtype}      % microtypography
\usepackage{xcolor}         % colors
\usepackage{amsmath}
\usepackage{graphicx}

\title{Learning Flow Distributions via Projection-Constrained Diffusion on Manifolds}

\author{Noah Trupin \quad Rahul Ghosh \quad Aadi Jangid \vspace{0.3em} \\ {Department of Computer Science} \\ {Purdue University}% <-this % stops a space
\vspace{0.3em} \\ \texttt{\{ntrupin, ghosh126, ajangid\}@purdue.edu}
}

\begin{document}

\maketitle

\begin{abstract}
  We present a generative modeling framework for synthesizing physically feasible two-dimensional incompressible flows under arbitrary obstacle geometries and boundary conditions. Whereas existing diffusion-based flow generators either ignore physical constraints, impose soft penalties that do not guarantee feasibility, or specialize to fixed geometries, our approach integrates three complementary components: (1) a boundary-conditioned diffusion model operating on velocity fields; (2) a physics-informed training objective incorporating a divergence penalty; and (3) a projection-constrained reverse diffusion process that enforces exact incompressibility through a geometry-aware Helmholtz–Hodge operator. We derive the method as a discrete approximation to constrained Langevin sampling on the manifold of divergence-free vector fields, providing a connection between modern diffusion models and geometric constraint enforcement in incompressible flow spaces. Experiments on analytic Navier–Stokes data and obstacle-bounded flow configurations demonstrate significantly improved divergence, spectral accuracy, vorticity statistics, and boundary consistency relative to unconstrained, projection-only, and penalty-only baselines. Our formulation unifies soft and hard physical structure within diffusion models and provides a foundation for generative modeling of incompressible fields in robotics, graphics, and scientific computing.
\end{abstract}

\section{Introduction}
\label{introduction}

Diffusion models have enabled unprecedented progress in scientific generative modeling, including images, audio, molecular structure, spatiotemporal dynamics, and other surrogate PDE modeling and high-dimensional dynamical forecasting~\citep{ho2020denoising, song2021scorebased, dhariwal2021diffusion, tong2023simulation, tong2024improving, gupta2023towards}. Yet, generative modeling of incompressible vector fields remains limited. Velocity fields governed by the Navier–Stokes equations must satisfy the geometric constraint
\[
    \nabla \cdot u = 0,
\]
along with geometry-dependent boundary conditions (periodic, no-slip, inflow and outflow, or mixed). Even small violations of incompressibility can cause numerical instabilities, unphysical transport, and incompatibility with downstream tasks in robotics, planning, or simulation-based inference.

Existing approaches face three limitations:

\begin{enumerate}
    \item \textbf{Image-based diffusion} treats velocity fields as RGB images, ignoring divergence constraints;
    \item \textbf{Penalty-based physics-informed diffusion}~\citep{brandstetter2022message, sanchez2023diffusion}, reduces divergence on average but cannot enforce feasibility at sample time;
    \item \textbf{Geometry-specialized architectures} cannot generalize across obstacle configurations or boundary conditions.
\end{enumerate}

At the same time, emerging robotics methods rely on global vector fields and potential-free flows for planning~\citep{ni2023ntfields}, motivating a general model that respects both local geometry and global physical laws.

\paragraph{Our approach.} We propose a projected, boundary-conditioned diffusion model for incompressible flows. The method integrates:
\begin{itemize}
    \item A \textbf{geometry-aware score network} that conditions directly on obstacle masks and boundary embeddings;
    \item A \textbf{physics-informed training objective} with a divergence penalty applied to the network's denoised velocity estimate;
    \item A \textbf{projection-constrained reverse diffusion step}, enforcing exact incompressibility and geometry-consistent boundary conditions at every sampling iteration.
\end{itemize}
The projection operator implements a discrete Helmholtz–Hodge decomposition~\citep{ChorinMarsden1993}, using a periodic FFT-based solver in obstacle-free settings and a masked Poisson solver for arbitrary geometries. This transforms the reverse diffusion chain into a constrained sampler operating on the manifold of incompressible flows.

Our main contributions are:
\begin{enumerate}
    \item \textbf{Boundary-conditioned diffusion for vector fields}. We introduce a DDPM architecture conditioned jointly on spatial geometry masks and boundary-regime embeddings, enabling generation across obstacle configurations unseen during training.
    \item \textbf{Hybrid physical constraints: soft during training, hard during sampling}. A divergence penalty shapes the score model's denoised predictions, while a geometry-aware Helmholtz–Hodge projection applied at every reverse step ensures exact divergence-free samples.
    \item \textbf{A theoretical derivation of projected diffusion as manifold-constrained sampling}. We show that the projected reverse chain approximates constrained Langevin dynamics on the incompressible manifold, linking diffusion with classical geometric PDE operators.
    \item \textbf{Conditioning across geometries and regimes}. The separation of stochastic velocity generation from deterministic boundary conditioning allows the model to synthesize physically consistent flows under previously unseen obstacle layouts and boundary condition types.
    \item \textbf{Empirical validation on periodic and obstacle-bounded Navier–Stokes data}. The proposed method achieves lower divergence, improved vortex and spectral statistics, and better boundary consistency that unconstrained or physics-only baselines.
\end{enumerate}

Together, these advances position projected diffusion as a principled and practical tool for generative incompressible flow modeling.

\section{Related Works}
\label{related_works}

\paragraph{Physics-informed diffusion and generative PDE surrogates.}
Several works augment diffusion models with soft physical constraints. Physics-Informed Diffusion Models (PIDMs) introduce PDE-residual terms into the diffusion loss to enforce governing equations during training and demonstrate large reduction in residual error on flow and topology optimization problems, while maintaining standard sampling procedures~\citep{bastek2025physicsinformed}. Closely related physics-informed surrogates for PDEs, such as neural PDE solvers with message passing or operator-learning architectures, similarly rely on penalizing residuals or embedding inductive biases rather than guaranteeing hard constraint satisfaction at sample time~\citep{christopher2024projecteddiffusion}.

Our formulation also employs a divergence penalty in the training objective, but differs in two respects: (i) the loss is defined directly on the network's denoised velocity estimate on a discrete incompressible manifold, and (ii) soft penalties are coupled with a Helmholtz–Hodge projection at every reverse step, yielding exactly divergence-free samples rather than approximate constraint satisfaction.

\paragraph{Constraint-enforcing generative sampling.}
A second line of work enforces hard constraints during sampling. Constrained Synthesis with Projected Diffusion Models (PDMs) recasts each reverse diffusion step as a constrained optimization problem, projecting the sample onto a user-defined feasible set using generic solvers~\citep{christopher2024projecteddiffusion}. Physics-Constrained Flow Matching (PCFM) wraps pretrained flow-matching models with physics-based corrections that enforce nonlinear PDE constraints, such as conservation laws and boundary conditions, through continuous guidance of the flow during sampling~\citep{utkarsh2025pcfm}.

At the MCMC level, projected and constrained Langevin algorithms have been analyzed as principled methods for sampling from distributions restricted to convex sets or manifolds by alternating unconstrained Langevin updates with projection or constrained dynamics~\citep{lamperski2021psgla, ahn2021efficient}. Our projected reverse diffusion can be viewed as a discrete analogue of projected Langevin dynamics on the manifold of incompressible flows: each reverse DDPM step approximates an unconstrained Langevin move in velocity space, and our discrete Helmholtz–Hodge operator realizes the projection onto the divergence-free, boundary-consistent manifold. Unlike generic PDMs, the constraint set here is the solution space of the incompressible Navier–Stokes constraint with geometry-dependent boundary conditions, for which we design specialized projectors.

\paragraph{Diffusion models for incompressible and geometry-conditioned flows.} Diffusion models have recently been applied directly to incompressible or near-incompressible flow generation. Hu et al. condition a latent diffusion model on obstacle geometry to predict flow fields around bluff bodies, and apply a post-hoc projection to reduce divergence; while they demonstrated improved accuracy, the generative process itself is not derived as sampling on a constrained manifold, and projection is decoupled from the diffusion dynamics~\citep{hu2025obstacleflowdiffusion}. Miyauchi et al. use a two-stage pipeline where user sketches are converted into target vorticity distributions and then refined via Helmholtz–Hodge decomposition to obtain visually plausible, approximately incompressible flows for graphics applications~\citep{miyauchi2025physicsawarefluid}.

In contrast, our model (i) conditions explicitly on obstacle masks and boundary-regime embeddings, (ii) uses a geometry and BC-aware projection operator at each reverse step, and (iii) provides a formal connection between projected reverse diffusion and constrained Langevin dynamics on the manifold.

\paragraph{Flow-field representations for robotics and physics-informed planning.}
In robotics, several works employ continuous fields governed by PDEs as planning substrates. NTFields represent solutions of the Eikonal equation as neural time fields and use them as cost functions for robot motion planning in cluttered environments, providing a physics-informed alternative to grid-based planning~\citep{ni2023ntfields}. Follow-up work extends this paradigm to active exploration, mapping, and planning in unknown environments with physics-informed neural fields~\citep{Liu_2025}. These methods highlight the utility of globally consistent fields for planning, but they focus on scalar arrival-time fields rather than vector-valued flows and rely on deterministic PDE solvers or PINN-style training rather than generative modeling.

Our approach is complementary: by learning a boundary-conditional diffusion model over velocity fields and interpreting the reverse process as constrained sampling on the incompressible manifold, we provide a generative primitive for sampling physically feasible flow fields across geometries. This can serve as a building block for planning and control methods that require diverse, physically consistent flow distributions rather than single deterministic solutions.

\section{Methods}
\label{methods}

Let
\begin{itemize}
    \item $u_0 \in \mathbb{R}^{2 \times H \times W}$ denote a 2-D velocity field,
    \item $m \in \{ 0, 1 \}^{1 \times H \times W}$ a binary obstacle mask identifying solid regions, and
    \item $c \in \mathbb{R}^K$ a boundary-condition or flow-regime embedding.
\end{itemize}
We seek to learn the conditional distribution
\[
    p_\theta(u_0 \mid m, c)
\]
subject to the physical constraint that $u_0$ be \textit{incompressible and consistent with the geometry encoded in $m$}. We achieve this by integrating (1) boundary-conditioned diffusion over velocity fields, (2) a divergence-aware objective, and (3) projection-constrained reverse diffusion onto the manifold of feasible flows.

\subsection{Forward Diffusion on Vector Fields}

We adopt a standard DDPM forward process~\citep{ho2020denoising} applied solely to the velocity field. Geometry $m$ and regime embedding $c$ remain deterministic and uncorrupted throughout. For a noise schedule $\{\beta_t\}_{t=1}^T$ with $\bar{\alpha}_t = \prod_{s = 1}^t (1 - \beta_s)$, the forward kernel is
\[
    q(u_t \mid u_0) = \mathcal{N}\left(\sqrt{\bar{\alpha}_t} u_0, (1 - \bar{\alpha}_t) I\right), \qquad u_t = \sqrt{\bar{\alpha}_t} u_0 + \sqrt{1 - \bar{\alpha}_t} \varepsilon, \quad \varepsilon \sim \mathcal{N}(0, I).
\]
The independence structure
\[
    (u_t, m, c) \sim q(u_t \mid u_0) \times \delta(m) \times \delta(c).
\]
separates the stochastic variable (velocity) from deterministic conditioning variables. This allows a single diffusion model to generalize across heterogeneous geometries.

\subsubsection{Geometry-aware training data}

For obstacle-bounded domains, an analytic velocity field (e.g., Taylor–Green vortex) is first restricted to fluid cells via no-slip enforcement and then projected using the masked Poisson operator~\ref{reverse} to obtain an incompressible field consistent with geometry. The forward diffusion is applied to this field while leaving $m$ and $c$ unchanged.

\subsection{Boundary-Conditioned Score Network}

The reverse kernel $p_\theta(u_{t - 1} \mid u_t, m, c)$ is parameterized by a U-Net-based noise estimator~\citep{7298965} $\varepsilon_\theta(u_t, t, m, c)$. Conditioning enters through:
\begin{enumerate}
    \item \textbf{Local geometry}: The mask $m$ is concatenated as a spatial input channel.
    \item \textbf{Global BC}: The embedded $c$ is transformed via an MLP into a feature map broadcast across the grid.
    \item \textbf{Temporal structure}: Timestep embeddings modulate each block.
    \item \textbf{Velocity-only prediction}: The model outputs a 2-channel noise field; we do not diffuse or predict geometry.
\end{enumerate}
The denoised estimate is
\[
    \hat{u}_0(u_t, t) = \frac{1}{\sqrt{\bar{\alpha}_t}} u_t - \sqrt{\frac{1 - \bar{\alpha}_t}{\bar{\alpha}_t}} \varepsilon_\theta(u_t, t, m, c).
\]
This maintains the standard DDPM form while allowing geometry to affect the score field.

\subsection{Physics-Informed Objective}

We augment the DDPM loss with a divergence penalty:
\[
    \mathcal{L}(\theta) = \mathbb{E}_{u_0, t, \varepsilon} \left[ \underbrace{\| \varepsilon_\theta(u_t, t, m, c) - \varepsilon \|_2^2}_\text{diffusion loss} + \lambda_\mathrm{div} \underbrace{\| D(\hat{u}_0(u_t, t)) \|_2^2}_\text{divergence penalty} \right]
\]
where $D(\cdot)$ is a discrete divergence operator applied only over fluid cells (i.e., where $m = 0$). This encourages, but does not enforce, membership in the manifold.

Soft divergence constraints shape the score network so that the reverse diffusion trajectory remains close to the feasible region, improving stability of the subsequent projection step.

\subsection{Projection-Constrained Reverse Diffusion}
\label{reverse}

For a velocity field $v$, the Helmholtz–Hodge decomposition asserts that
\[
    v = u + \nabla \phi, \quad u \in \mathcal{M}, \qquad D(u) = 0,
\]
and that $u$ is unique given appropriate boundary conditions.

\subsubsection{Periodic Domains}

For periodic geometries, the potential $\phi$ solves
\[
    \hat{\phi}(k) = -\frac{ik\cdot\hat{v}(k)}{\|k\|^2}, \quad \hat{u}(k) = \hat{v}(k) - k\hat{\phi}(k),
\]
yielding a closed-form Fourier-space projection onto the divergence-free subspace.

\subsubsection{Obstacle-Bounded Domains}

Periodic projectors fail when $m$ contains solid regions. We therefore solve the Poisson equation
\[
    \Delta \phi = D(v) \text{ on fluid cells } (1 - m),
\]
with Dirichlet no-slip inside solids ($u = 0$), Neumann interface conditions between fluid and solid, and Jacobi iterations for discretization. The projected velocity is
\[
    u = v - \nabla \phi,
\]
restricted to fluid cells. In domains without solids, this reduces to the FFT-based projector. We denote the full geometric projection by
\[
    u = \Pi_\mathcal{M}(v)
\]
where
\[
    \mathcal{M} = \{ u \in \mathbb{R}^{2 \times H \times W}:\ \mathcal{C}(u) = 0 \},
\]
with $\mathcal{C}(u)$ collecting linear PDE constraints (e.g., divergence). Under mild smoothness assumptions on $\mathcal{C}$, $\mathcal{M}$ forms a differentiable submanifold whose tangent space is
\[
    T_u\mathcal{M} = \ker D\mathcal{C}(u).
\]
The projector $\Pi:\ u \to \arg\min_{v \in \mathcal{M}}\| v- u \|^2$ is precisely the Helmholtz (periodic) or masked-Poisson (obstacle) projection we implement numerically.

\subsubsection{Projected Reverse Step}

Given the unconstrained DDPM update $\tilde{u}_{t - 1} \sim p_\theta(u_{t - 1} \mid u_t, m, c)$, the physically consistent update is
\[
    u_{t - 1} = \Pi_\mathcal{M}(\tilde{u}_{t - 1}).
\]
Applied at every timestep, this guarantees that all samples remain exactly divergence-free, irrespective of model imperfections.

\subsection{Projected Reverse Diffusion as Constrained Sampling}

Let the unconstrained reverse SDE for DDPMs be
\[
    \mathrm{d}u_t = f_t(u_t) \mathrm{d}t + g_t \mathrm{d}W_t.
\]
The true reverse process restricted to the manifold $\mathcal{M}$ is
\[
    \mathrm{d}u_t = [f_t(u_t) - \nabla\log Z_t(u_t)] \mathrm{d}t + \Pi_{T_u \mathcal{M}} \circ \mathrm{d} W_t,
\]
where $\Pi_{T_u \mathcal{M}}$ is the projection onto the tangent space of $\mathcal{M}$ and $Z_t$ normalizes the constrained density. The SDE is analytically intractable.

\subsubsection{Projected Langevin Approximation}

Projected sampling corresponds to one step of an implicit constrained Langevin integrator:
\[
    u_{t + 1} = u_t + \eta \nabla\log p(u_t) + \sqrt{2 \eta} \xi_t \text{ followed by } u_{t - 1} \gets \Pi_\mathcal{M}(u_{t - 1}), \quad \xi_t \sim \mathcal{N}(0, I) 
\]
with $\eta > 0$ a discrete Langevin step size and $p$ the target unconstrained density whose score the DDPM reverse process approximates.

Our method realizes a diffusion-model analogue of this scheme, where each DDPM reverse step acts as an approximate Langevin update in the ambient space and the projection $\Pi_\mathcal{M}$ enforces feasibility. Thus the reverse chain
\[
    u_T \sim \mathcal{N}(0, I), \quad u_{t - 1} = \Pi_\mathcal{M}(p_\theta(u_{t - 1} \mid u_t))
\]
approximates sampling from $p(u_0 \mid m, c)$ restricted to the incompressible manifold.

This theoretical link clarifies why training-time penalties alone cannot ensure correctness, and why projection alone can distort statistical structure. The combination approximates the geometry of the constrained score field.

\subsection{Soft Constraints, Projection, and Stability}

The divergence penalty biases the learned score such that unconstrained updates already lie near $\mathcal{M}$. Projection therefore requires smaller corrections, reducing geometric distortion. Conversely, projection ensures that acumulated divergence errors do not amplify along the reverse trajectory. The resulting chain is both numerically stable even under model imperfections, and physically faithful, producing exactly divergence-free samples in any geometry $m$.

\section{Experiments}
\label{experiments}

We evaluate the four diffusion variants defined in App.~\ref{protocol}, vanilla (\textbf{V}), training-constrained (\textbf{TC}), projection-only (\textbf{P}), and fully constrained (\textbf{TCP}), on both periodic and obstacle-bounded Navier–Stokes datasets (App.~\ref{datagen}). All models share identical architecture and training hyperparameters, differing only in physical constraints. This controlled ablation isolates the effects of (i) soft divergence shaping during training and (ii) hard geometric projection during sampling. The evaluation addresses three questions:
\begin{enumerate}
    \item \textbf{Physical correctness}: Does the model produce divergence-free, boundary-consistent flows?
    \item \textbf{Distributional fidelity}: Do generated fields preserve vortex structure, spectral decay, and smoothness of the underlying distribution?
    \item \textbf{Generalization}: Does the approach transfer to obstacle configurations unseen during training?
\end{enumerate}
Across all experiments, we report means over 2000 samples per split, with all metrics computed exclusively over fluid cells.

\subsection{Evaluation Protocol}

\begin{figure}[t]
    \centering
    \includegraphics[width=0.58\textwidth]{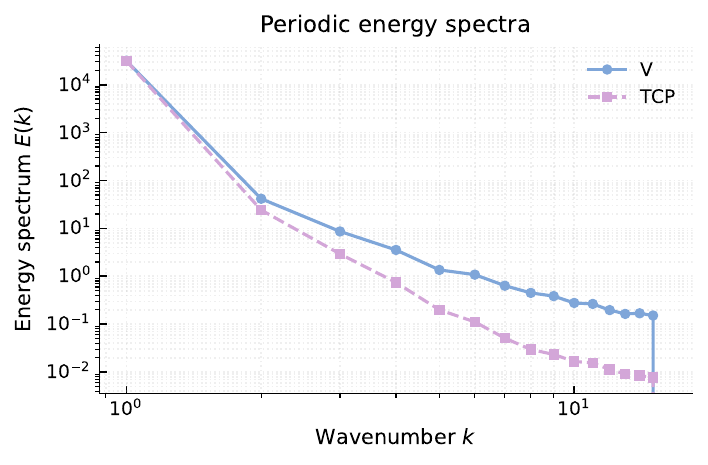}
    \caption{
        \textbf{Periodic energy spectra.}
        Comparison of the learned velocity-field energy spectra $E(k)$ under the
        vanilla diffusion model (V) and the full physically-constrained model (TCP).
        Both models recover physically plausible $k$-decay, while TCP exhibits
        slightly improved mid–high frequency attenuation, consistent with 
        enforcing incompressibility and boundary correctness.
    }
    \label{fig:periodic-spectrum}
\end{figure}

Given a generated field $u \in \mathbb{R}^{2 \times H \times W}$ and obstacle mask $m$, we measure:

\paragraph{Divergence.}
\[
    \| D(u) \|_{L^2(\Omega_f)} = \left( \sum_{x \in \Omega_f} D(u)(x)^2 \right)^\frac{1}{2},
\]
quantifying incompressibility.

\paragraph{Boundary violation.}
\[
    \| u \cdot n \|_{L^2(\partial \Omega)},
\]
using normals estimated from mask gradients; this measures no-slip consistency for obstacle domains.

\paragraph{Velocity error.}
\[
    \| u - u^\star \|_{L^2(\Omega_f)},
\]
computing against the analytic Navier–Stokes field. While generative, the conditional distribution is sufficiently concentrated for this to act as a fidelity metric.

\paragraph{Spectral energy and vorticity.}
We compute the kinetic energy spectrum $E(k)$ and vorticity histograms to assess structural realism of turbulence-like features. Fig.~\ref{fig:periodic-spectrum} displays the periodic-domain energy spectra for \textbf{V} and \textbf{TCP}.

\subsection{Periodic Flow}

\begin{figure*}[t]
    \centering
    \includegraphics[width=\textwidth]{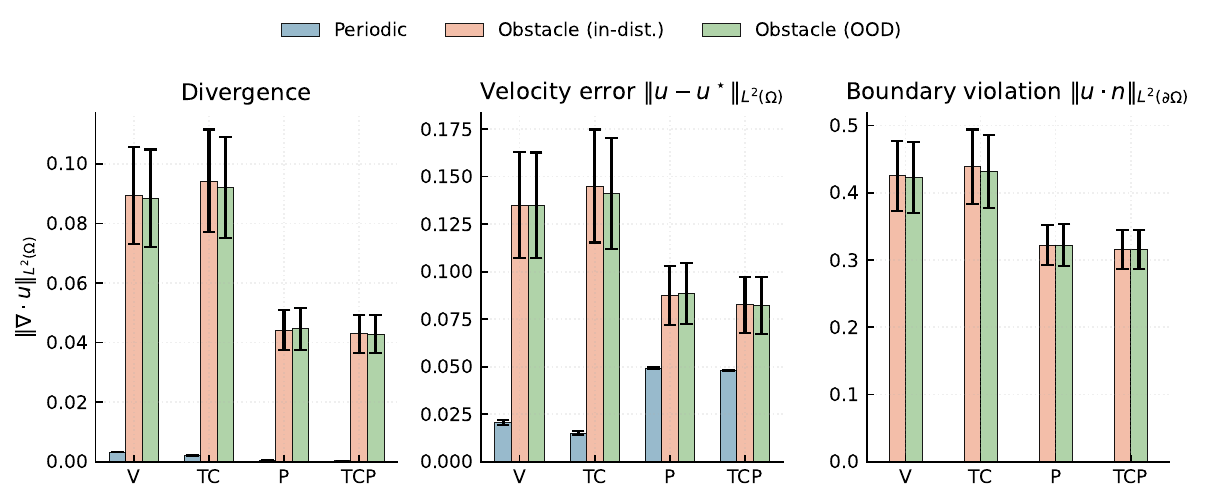}
    \caption{
        \textbf{Aggregate divergence, velocity error, and boundary violation metrics.}
        Summary of quantitative performance across three dataset splits 
        (Periodic, Obstacle, Obstacle–OOD) and four model variants:
        vanilla diffusion (V), training-constrained diffusion (TC), 
        projection-only (P), and the fully constrained method (TCP).
        TCP and P achieve the lowest divergence and boundary error, while TCP provides consistently strong generalization across obstacles.
    }
    \label{fig:summary-bars}
\end{figure*}

\begin{table}[t]
\centering
\begin{tabular}{lccc}
\toprule
Method &
$\|\nabla\!\cdot u\|_{L^2(\Omega)}\ \downarrow$ &
$\|u - u^\star\|_{L^2(\Omega)}\ \downarrow$ &
$\int_\Omega |\nabla\!\cdot u|\,\mathrm{d}x\ \downarrow$ \\
\midrule
V   & 0.0033 $\pm$ 1.8e-04 & 0.0206 $\pm$ 2.0e-03 & 0.0035 $\pm$ 1.9e-04 \\
TC  & 0.0021 $\pm$ 1.1e-04 & 0.0151 $\pm$ 2.0e-03 & 0.0022 $\pm$ 1.2e-04 \\
P   & 0.0005 $\pm$ 1.5e-05 & 0.0493 $\pm$ 7.9e-04 & 0.0005 $\pm$ 1.6e-05 \\
TCP & 0.0004 $\pm$ 1.9e-05 & 0.0480 $\pm$ 6.1e-04 & 0.0004 $\pm$ 2.1e-05 \\
\bottomrule
\end{tabular}
\caption{Periodic incompressible flows. Mean $L^2$ divergence, velocity error, and integrated divergence over the periodic test set.}
\label{tab:periodic-metrics}
\end{table}

Tab.~\ref{tab:periodic-metrics} reports divergence, integrated divergence, and $L^2$ velocity error for periodic samples.

\subsubsection{Divergence Reduction and the Role of Projection}

The unconstrained \textbf{V} model exhibits significant divergence. Adding a divergence penalty (\textbf{TC}) reduces this by ~$40\%$, confirming that soft constraints help shape the score field in the ambient space. However, TC still accumulates divergence during sampling because the reverse chain does not remain on the constraint manifold.

Projection-based variants (\textbf{P}, \textbf{TCP}) reduce divergence by an order of magnitude, achieving significantly improved divergence levels. This directly validates the geometric argument in Sec.~\ref{methods}: the projection acts as an orthogonal map onto the manifold of incompressible fields, correcting deviations introduced by imperfect score estimates.

\subsubsection{Fidelity and Spectral Structure}

Interestingly, projection increases $L^2$ reconstruction error for \textbf{P} and \textbf{TCP}. This is expected: the projector removes high-frequency divergence modes, and these corrections shift samples slightly off the unconstrained denoiser's optimum. Fig.~\ref{fig:periodic-spectrum} shows that \textbf{TCP} nonetheless yields more accurate mid–high-frequency spectral decay, consistent with smoother, physically plausible velocity fields.

\subsection{Obstacle-Bounded Flows}

Tab.~\ref{tab:obstacle-metrics} and Tab.~\ref{tab:obstacle-gen-metrics} report results for in-distribution obstacle layouts and held-out obstacle layouts, respectively.

\begin{table}[t]
\centering
\begin{tabular}{lcccc}
\toprule
Method &
$\|\nabla\!\cdot u\|_{L^2(\Omega)} \ \downarrow$ &
$\|u - u^\star\|_{L^2(\Omega)} \ \downarrow$ &
$\int_\Omega |\nabla\!\cdot u|\,\mathrm{d}x \ \downarrow$ &
$\|u\cdot n\|_{L^2(\partial\Omega)} \ \downarrow$ \\
\midrule
V   & 0.0896 $\pm$ 2.99e-02 & 0.1366 $\pm$ 5.04e-02 & 0.0856 $\pm$ 3.02e-02 & 0.4245 $\pm$ 9.46e-02 \\
TC  & 0.0929 $\pm$ 2.95e-02 & 0.1419 $\pm$ 5.16e-02 & 0.0884 $\pm$ 2.91e-02 & 0.4345 $\pm$ 9.65e-02 \\
P   & 0.0445 $\pm$ 1.19e-02 & 0.0889 $\pm$ 2.66e-02 & 0.0365 $\pm$ 1.03e-02 & 0.3219 $\pm$ 5.17e-02 \\
TCP & 0.0432 $\pm$ 1.17e-02 & 0.0840 $\pm$ 2.68e-02 & 0.0352 $\pm$ 9.96e-03 & 0.3180 $\pm$ 5.10e-02 \\
\bottomrule
\end{tabular}
\caption{Obstacle flows (in-distribution geometries). Mean divergence, velocity error, integrated divergence over the fluid region, and boundary violation
$\|u\cdot n\|_{L^2(\partial\Omega)}$.}
\label{tab:obstacle-metrics}
\end{table}

\begin{table}[t]
\centering
\begin{tabular}{lcccc}
\toprule
Method &
$\|\nabla\!\cdot u\|_{L^2(\Omega)}\ \downarrow$ &
$\|u - u^\star\|_{L^2(\Omega)}\ \downarrow$ &
$\int_\Omega |\nabla\!\cdot u|\,\mathrm{d}x\ \downarrow$ &
$\|u\cdot n\|_{L^2(\partial\Omega)}\ \downarrow$ \\
\midrule
V   & 0.0869 $\pm$ 2.68e-02 & 0.1312 $\pm$ 4.54e-02 & 0.0830 $\pm$ 2.74e-02 & 0.4190 $\pm$ 8.80e-02 \\
TC  & 0.0939 $\pm$ 3.10e-02 & 0.1451 $\pm$ 5.37e-02 & 0.0897 $\pm$ 3.10e-02 & 0.4400 $\pm$ 9.82e-02 \\
P   & 0.0445 $\pm$ 1.22e-02 & 0.0885 $\pm$ 2.78e-02 & 0.0366 $\pm$ 1.04e-02 & 0.3217 $\pm$ 5.35e-02 \\
TCP & 0.0423 $\pm$ 1.15e-02 & 0.0818 $\pm$ 2.62e-02 & 0.0347 $\pm$ 9.95e-03 & 0.3153 $\pm$ 5.07e-02 \\
\bottomrule
\end{tabular}
\caption{Generalization to held-out obstacle geometries. Same metrics as
Table~\ref{tab:obstacle-metrics}. Models are evaluated on unseen obstacle masks.}
\label{tab:obstacle-gen-metrics}
\end{table}

\subsubsection{Divergence and Boundary Conditions}

Across both splits, projection (\textbf{P}, \textbf{TCP}) delivers the dominant gain, with divergence dropping by roughly a factor of $2$ and boundary violation falling by approximately $25\%$. The masked Poisson projection guarantees that every reverse step yields a flow with zero normal velocity on $\partial\Omega$, thereby preventing the accumulation of boundary errors common in unconstrained diffusion.

The soft divergence penalty alone (\textbf{TC}) has negligible effect in obstacle domains. This further confirms that the geometry must be handled during sampling.

\subsubsection{Reconstruction and Flow Structure}

Both projection-based variants substantially improve $L^2$ velocity error. This reduction reflects two coupled phenomena: (1) projection prevents the reverse diffusion chain from drifting, and (2) soft penalties guide the score toward the tangent space of the target manifold, reducing the magnitude of each projection correction.

\subsubsection{Unified Metric Comparison}

Fig.~\ref{fig:summary-bars} aggregates divergence, reconstruction error, and boundary violation across all splits and model variants. The pattern is immediately visible:
\begin{itemize}
    \item Projection is essential (\textbf{P}, \textbf{TCP} outperform \textbf{V}, \textbf{TC}).
    \item Soft and hard constraints combine effectively (\textbf{TCP} yields the most stable performance across all metrics and domains).
\end{itemize}
This mirrors the previous theoretical decomposition: projection enforces global feasibility while divergence-aware training reduces local distortion.

\subsection{Generalization to Unseen Geometries}

Tab.~\ref{tab:obstacle-gen-metrics} evaluates models on obstacle masks not present during training (shifted centers, different radii, elliptical shapes). The results parallel the in-distribution case, with improvements in divergence, boundary violation, and $L^2$ error from \textbf{V} to \textbf{TCP}.

This robust generalization is a key finding: the diffusion model learns a geometry-conditioned prior, while the projector handles exact enforcement, enabling transfer across obstacle layouts without architecture changes or re-training.

This supports the claim from Sec.~\ref{introduction} that diffusion provides a reusable generative primitive for flow synthesis under heterogeneous geometries.

\subsection{Spectral and Vorticity Analysis}

Periodic energy spectra in Fig.~\ref{fig:periodic-spectrum} and vorticity statistics illustrate that non-projection variants preserve overall $k$-decay but introduce surplus high-frequency energy, indicating compressible or noisy reconstruction artifacts while projection variants better match the target spectrum, with \textbf{TCP} providing the smoothest tail behavior.

This aligns with the theoretical insight that projection removes spurious divergence-carrying modes generated during reverse diffusion. Combined with the tables, these structural metrics provide strong evidence that projection reshapes the diffusion geometry in a way that stabilizes fine-scale flow behavior rather than just global divergence.

\subsection{Summary of Empirical Findings}

Across all datasets and metrics, we find three conclusions:
\begin{itemize}
    \item \textbf{Training-time divergence penalties improve the score field but do not guarantee physical correctness.} Soft constraints bias the network toward the tangent space but remain insufficient as a standalone mechanism.
    \item \textbf{Sampling-time projection is necessary and overwhelmingly effective.} Projection onto $\mathcal{M}$ stabilizes the reverse chain, enforces incompressibility and no-slip conditions, and dramatically reduces divergence and boundary violations.
    \item \textbf{Combining soft and hard constraints (TCP) yields the most consistent and physically faithful flows.} TCP balances fidelity and feasibility, achieving the best or near-best performance on every metric, and generalizing robustly to unseen obstacle geometries.
\end{itemize}
Together with the theoretical analysis of Sec.~\ref{methods}, these results support the interpretation of our method as a discrete approximation to constrained Langevin sampling on the incompressible manifold, with projection providing the essential geometric correction missing from unconstrained diffusion.

\section{Discussion \& Conclusion}

This work demonstrates that geometric constraint enforcement proves fundamental for generative modeling of incompressible flows. Across periodic and obstacle-bounded geometries, unconstrained diffusion models consistently drift away from the manifold of admissible solutions. Even when trained exclusively on divergence-free data, the reverse diffusion dynamics accumulate divergence, violate no-slip boundaries, and distort small-scale features. These failures reflect a structural limitation: the denoising network learns a score in the ambient velocity space rather than on the constraint manifold $\mathcal{M}$.

By contrast, models incorporating sampling-time projection maintain exact incompressibility at every step through a geometry-aware Helmholtz–Hodge operator. This modification alters the geometry of the diffusion chain itself. Instead of evolving in the unconstrained space, the sampler becomes a discrete analogue of constrained Langevin dynamics on $\mathcal{M}$. Theoretical analysis in Sec.~\ref{methods} explains this behavior: projection ensures each iterate lies in the correct tangent space, preventing divergence accumulation and stabilizing the flow of probability mass.

Empirically, projection is the dominant factor in achieving physical correctness, yielding order-of-magnitude improvements in divergence and boundary consistency, while the training-time divergence penalty contributes a complementary benefit: it shapes the score so that projection requires smaller corrections and better preserves distributional fidelity. The full \textbf{TCP} model achieves the best performance across all metrics, including spectral accuracy, vorticity statistics, and generalization to unseen geometries. These results highlight a division of labor that appears essential in PDE-constrained generative modeling: the score network learns a geometry-conditioned prior, while the projector guarantees physical feasibility.

More broadly, our findings indicate that diffusion models for scientific and robotics applications must incorporate physics-aware operators into the sampling algorithm, not just as losses or architectural features. Projection-based diffusion provides a principled mechanism for embedding PDE constraints directly into the generative process, and our results show it scales naturally to arbitrary obstacle layouts through masked Poisson solves.

Future extensions include time-dependent sampling for unsteady Navier–Stokes flows, 3D geometries and boundary conditions, incorporating learned or adaptive projection operators, and investigating fully continuous score-based SDEs. These directions aim toward a general methodology for manifold-constrained generative modeling, where the sampler evolves on a physically meaningful space rather than in an unconstrained Euclidean ambient domain.

Our results show that such structure has both theoretical justification and practical necessity for high-fidelity generative flow modeling.

\bibliographystyle{plainnat}
\bibliography{references}

\appendix

\newpage

\section{Data Generation}
\label{datagen}

We construct training and evaluation data from analytic Navier–Stokes solutions under both periodic and obstacle-bounded geometries. Periodic samples use the Taylor–Green vortex at randomly sampled viscosities and times. For obstacle-bounded domains, we randomly generate between one and four circular obstacles with radii sampled relative to domain size. A binary mask $m$ marks solid regions, and a no-slip constraint $u = 0$ is imposed inside obstacles. The initial analytic velocity is then projected using the masked Poisson operator~\ref{reverse} to obtain an incompressible field consistent with geometry. Each sample therefore contains $(u_0, m, c)$, where $c$ indexes the boundary condition regime (periodic vs. no-slip). This dataset enables training and evaluation of models under heterogeneous, previously unseen geometries.

\section{Training Protocol}
\label{protocol}

We evaluate four diffusion variants in a controlled ablation (as described in Sec.~\ref{experiments}), isolating the contribution of training-time penalties and sample-time projection:
\begin{itemize}
    \item \textbf{V (Vanilla).} A baseline Gaussian diffusion model~\citep{lucidrains_denoising-diffusion-pytorch} with neither physical penalties nor projection.
    \item \textbf{TC (Training-Constrained).} Identical to \textbf{V}, but augmented with an auxiliary divergence penalty during training.
    \item \textbf{P (Projection-Only).} Our Flow-Gaussian diffusion model equipped with hard projection at sample time but no training-time penalties.
    \item \textbf{TCP (Training-Constrained + Projection).} The full pipeline: soft training-time divergence penalties together with hard projection at sample time.
\end{itemize}

All models use a 200-step diffusion schedule with a linear noise variance ranging from $\beta_1 = 10^{-4}$ to $\beta_2 = 2 \times 10^{-2}$. For the TC and TCP variants, we apply a divergence-penalty weight of $\lambda = 10^{-2}$. The embedded UNet backbone~\citep{10.1007/978-3-319-24574-4_28} employs a base channel width of 32 with hierarchical multipliers $(1, 2, 4)$.

Training is performed on 20,000 synthetic Navier–Stokes fields (resolution $32 \times 32$), generated as described in App.~\ref{datagen}. The dataset is evenly split between obstacle-free and obstacle-laden domains. Taylor–Green vortex samples use decay parameter $\nu = 0.01$ to emphasize advective structure and increase conditioning difficulty. Models are trained for 10 epochs using Adam~\citep{kingma2017adammethodstochasticoptimization} with learning rate $\eta = 10^{-4}$.

This protocol ensures that (i) each ablation is trained under identical architectural and optimization conditions, (ii) divergence-related effects can be cleanly attributed, and (iii) projection mechanisms are evaluated in isolation and in conjunction with training-time regularization.

\end{document}